# Modelling and Predicting the Conditional Variance of Bitcoin Daily Returns: Comparsion of Markov Switching GARCH and SV Models

Dennis Koch[*] and Vahidin Jeleskovic[**] and Zahid I.Younas[***]

**Abstract**

This paper introduces a unique and valuable research design aimed at analyzing Bitcoin price volatility. To achieve this, a range of models from the Markov Switching-GARCH and Stochastic Autoregressive Volatility (SARV) model classes are considered and their out-of-sample forecasting performance is thoroughly examined. The paper provides insights into the rationale behind the recommendation for a two-stage estimation approach, emphasizing the separate estimation of coefficients in the mean and variance equations.

The results presented in this paper indicate that Stochastic Volatility models, particularly SARV models, outperform MS-GARCH models in forecasting Bitcoin price volatility. Moreover, the study suggests that in certain situations, persistent simple GARCH models may even outperform Markov-Switching GARCH models in predicting the variance of Bitcoin log returns.

These findings offer valuable guidance for risk management experts, highlighting the potential advantages of SARV models in managing and forecasting Bitcoin price volatility.

[*] Independent researcher, Germany. Email: dennis.herle@web.de
[**] Submitting author, Humboldt-Universität zu Berlin, Germany. Email: vahidin.jeleskovic@hu-berlin.de
[***] Berlin School of Business and Innovation, Germany. Email: zahid1132_gcu@yahoo.com

# 1. Introduction

In abstract terms, cryptocurrencies, coins, and tokens are unique digital resources of an autarkic functioning network. The network's resources, which are merely interpreted as currencies, are based on cryptography and cannot be duplicated. Each network of a certain cryptocurrency is characterized by a theoretically immutable distributed ledger (the so-called blockchain) that contains the entire transaction history of the network's participants and a bookkeeping and resource distribution mechanism which is based on fundamental concepts of computer science and cryptographic algorithms (Antonopoulos, 2017).

Due to the multifaceted nature of cryptocurrencies and the potential future demand, investigating the fluctuations of cryptocurrency returns in order to adjust risk management systems could be valuable. Traditional volatility prediction methods are presumably not applicable for cryptocurrencies. For example, a part of the recent research on the volatility of Bitcoin returns indicates that the application of generalized autoregressive conditional heteroscedasticity (GARCH) based models to describe the conditional variance of Bitcoin returns is problematic. The main problem lies in the frequently estimated non-stationarity of the variance equation. Non-stationarity occurs when the estimated sum of ARCH and GARCH parameters of the conditional variance equation are larger than one. The theoretical requirements for variance stationarity are unfulfilled in these cases. A prediction based on these models generates exploding processes which could be highly impractical for risk management considerations. Another problem is the identification of a particular GARCH model type. The period in which the GARCH models are estimated influences which of the GARCH models is the best fit according to the common information criteria, Akaike (AIC), Bayes-Schwarz (BIC) and Hannan-Quinn (HQ). The dependence of the best GARCH model type on the period could be an indication for structural changes in the underlying data-generating process. The second section of this paper reviews previous research on GARCH models estimated on cryptocurrency returns. The focus will be on Bitcoin, as it is the oldest and, according to market capitalization, the most important cryptocurrency.

In light of the limitations observed in previous literature (Caporale et al., 2003; Bouri et al., 2017; Dyhrberg, 2016; Kurihara and Fukushima, 2018; Katsiampa, 2017; Guo, 2022; Dudek et al., 2023), this research makes a unique contribution to the field of cryptocurrencies and financial risk management. It introduces readers to the methodology of Markov Switching-GARCH (MS-GARCH) and Stochastic Autoregressive Volatility (SARV) models. MS-GARCH models address issues related to possible structural changes and non-stationarity by assuming the existence of multiple data-generating processes (regimes) concurrently. The actual realizations of the conditional variance can then switch from one data-generating process to another one in the next period – or alternatively stay in the actual regime – with a certain probability. Compared to GARCH models, these specifications allow the conditional variance of returns to be explained by using multiple variance equations instead of just one. The SARV models used in this paper are based on the well-known Kalman filter. A special characteristic of this algorithm is the dynamic (real time) extraction of latent values from measurements. For example, the conditional variance of a return series can be seen as a latent (unobservable) variable. The very heart of the algorithm's extraction process is the adjustment of unconditional expectations about a certain variable (like the variance) by an information weight, the so-called Kalman gain. The dynamic adjustments of the Kalman gain could lead to superior variance predictions by SARV models (Bouri et al.,2017; Kurihara and Fukushima; 2018).

Dyhrberg (2016) suggests that the dynamics of the Bitcoin price should be simultaneously modeled in both the conditional first moment (mean equation) and the conditional second moment (variance equation). We employ a two-step estimation procedure, where the mean equation is estimated first, followed by the application of GARCH models and SARV models to the estimated residuals. There are several reasons for this approach. First, Engle (1982) points out that the two-step estimation procedure leads to consistent estimates for both the coefficients in the mean equation as well as in the variance equation, essentially resembling the FGLS procedure. Second, optimizing model specifications becomes significantly easier for researchers when they want to employ flexible lag structures and model specifications in both equations. For example, if we have 6 variables in the mean equation and 4 variables in the variance equation, there are a total of $\sum_{j=0}^{10}\binom{10}{j} = 1024$ possible model specifications with at least an intercept in both equations. When researchers first optimize the mean equation, there are a total of 64 possible specifications, and subsequently, in the variance equation, there are 16 possible specifications, greatly simplifying the process of determining the optimal model specification. A third important reason is that numerical optimization works much better in the two-step procedure, and convergence in numerical optimization is easier to achieve. Finally, it is easier and more intuitive to compare volatility models when they are based on the same mean equation, which is considered to be the best specification for the conditional mean.

In the empirical part of this paper, which is covered in section four, various MS-GARCH and SARV models are fitted to the log returns of the U.S. dollar per Bitcoin exchange rate (USD/BTC). These models are estimated after coefficients in the conditional mean equation have been determined. The best-fitting MS-GARCH models, as determined by the AIC, BIC, and HQ information criteria, are compared with the best-fitting SARV models, which are selected based on a least squares target function. The comparison is based on their respective abilities to predict the variance of Bitcoin log returns. A one-step-ahead variance prediction is performed repeatedly for the out-of-sample period, with model parameters being updated daily for each subsequent one-step-ahead prediction. In other words, the model parameters are refreshed on a daily basis. The prediction quality of MS-GARCH and SARV models is assessed using common error measures such as mean absolute error (MAE), mean squared error (MSE), and quasi-likelihood (QL). According to the QL loss measure, it is found that SARV models outperform MS-GARCH models in terms of their ability to predict variance. Furthermore, the paper suggests that a simple GARCH model with stationary variance and persistence close to unity (the sum of the parameters in the conditional variance equation) could serve as a second-best solution for risk management purposes. The final section of the paper provides a conclusion of the results and offers insights into potential avenues for future research.

## 2. Literature Review

Many researchers estimated and presented explosive variance processes, often without further discussion on this issue. The research extends the research of Caporale et al. (2003) and points out that, even if stationary processes were estimated, the high persistence of the variance, which was often near one (as is the case for Integrated-GARCH models), is a strong indicator for structural changes of the underlying data-generating process (Caporale et al., 2003), this means that the estimation of I-GARCH models as one of the best suitable models could be a consequence of structural changes. Caporale et al. (2003) demonstrated on the basis of a Monte-Carlo simulation that the estimation of an I-GARCH model can originate from an erroneous model specification[II]. The true stochastic process in the study of Caporale et al. (2003) followed an MS-GARCH model with two regimens and an order of one for each of the GARCH and ARCH terms in the conditional variance equation. Therefore, the study asserts that MS-GARCH models could be a solution to overcome non-stationarity and can be used for prediction purposes.

In previous literature from Kurihara and Fukushima (2018) and Katsiampa (2017), the best model judged by the AIC, BIC, and HQ to describe the variance of Bitcoin returns is a C-GARCH model. Additionally, both studies report the estimated parameters of the alternative GARCH models. The sum of the reported parameters shows that the simple GARCH model does not fulfil the condition of variance stationarity in both studies. In addition, the reported T-

---

[II] The I-GARCH model is characterized by the sum of ARCH and GARCH parameters of a simple GARCH model being close to one. Despite their covariance non-stationarity, I-GARCH models are considered as strictly stationary models in the literature (Bollerslev 2008, p. 18).

GARCH model in Katsiampa (2017) can also be classified as non-stationary[III]. Considering that Kurihara and Fukushima (2018) and Katsiampa (2017) have chosen largely overlapping periods and analyzed index data, the almost identical results are not surprising. Similarly, the results from Bouri et al. (2017) imply stationarity of the estimated GJR-GARCH models. Considering the high persistence of the variance process and the insignificance of the asymmetry terms, non-stationarity cannot be completely denied, at least for the entire and second analysis period. Furthermore, the equation for the conditional mean strongly depends on the analyzed period. In the second period, for example, Bouri et al. (2017,) reported a conditional mean equation without any autocorrelation. The conditional mean is a autoregressive moving average process with an order of zero (ARMA(0,0)) with a constant. Further, Chu et al. (2017) identified an I-GARCH model as the best model to describe the conditional variance of Bitcoin returns, taking into account a total of five information criteria. One noteworthy aspect is that all evaluated information criteria suggest the I-GARCH model to be the best model to describe the variance of Bitcoin returns. By referring to the paper by Caporale et al. (2003), Chu et al. (2017, p. 12) noted in the conclusion of their paper that their results might be compromised by structural changes of the underlying data-generating process.

The best GARCH models reported by Bouoiyour and Selmi (2016) are presumably the most inappropriate GARCH models the authors estimated. The information criteria reported in the appendix of their work suggest different GARCH model variations than those discussed in the text. Since the authors do not provide estimated parameters of the considered GARCH models, a definitive conclusion cannot be drawn about the stationarity of individual GARCH models. However, the fact that the information criteria suggest different GARCH models is remarkable. These change from the first to the second analyzed period. Thus, the AIC information criterion, for example, suggests a T-GARCH model[IV] in the first period and an E-GARCH model in the second. The abrupt change of the models classified as best could be seen as an indicator of structural changes, at least when it comes to capturing the asymmetric behavior of Bitcoin returns.

According to Bouri et al. (2017) and Dyhrberg (2016), only one GARCH model specification is used: GJR-GARCH. The analysis period in Dyhrberg (2016, p. 140) lies within the analysis period of Bouri et al. (2017, p. 3). However, the equations for the conditional mean of Bitcoin

---

[III] Unfortunately, Katsiampa (2017, p. 3) is not precise about the correct naming of the estimated "T-GARCH" model. She refers to the study from Bouoiyour and Selmi (2016) as well as to the study of Bouri et al. (2017), which both estimated "T-GARCH" models. In Bouri et al. (2017, p. 6), the GJR-GARCH model by Glosten et al. (1993) is used to map asymmetry, whereas in Bouoiyour and Selmi (2016, p. 10) the T-GARCH model from Zakoian (1994) is used. According to the reported "T-GARCH" variance equation, Katsiampa (2017, p. 5) estimated a GJR-GARCH model. The GJR-GARCH (1,1) model's condition for stationarity under the assumption of normally distributed innovations is given by $\alpha_1 + \beta + 0.5\gamma < 1$, where $\gamma$ is called the asymmetry term in the variance equation given by $\sigma_t^2 = \alpha_0 + \alpha_1 Z_{t-1}^2 + \beta \sigma_{t-1}^2 + \gamma Z_{t-1}^2 I_{Z_{t-1}<1}$. $I_{Z_{t-1}<1}$ is an indicator function that takes on one if the innovations realization is negative (Bouri et al. 2017, p. 6).
[IV] This refers to the threshold GARCH model from Zakoian (1994).

returns differ from those in Bouri et al. (2017, p. 8). Due to Dyhrberg's (2016, p. 140) research question (whether Bitcoin is suitable as a hedge against exchange rate fluctuations in the USD/EUR or USD/GBP exchange rate and fluctuations of the FTSE stock index), the author includes the returns of the exchange rates and the stock index in the conditional mean equation of Bitcoin returns. Dyhrberg (2016) has a total of three models (the "USD/EUR model", "USD/GBP model", and the "FTSE model") for the conditional mean equation of Bitcoin returns in which the logarithmic price of Bitcoin at lag one as well as the current return and the return at lag one of a particular exchange rate (USD/EUR or USD/GBP) or the stock index (FTSE) are included. In every conditional mean specification made by Dyhrberg (2016, p. 142), the GJR-GARCH model is intended to describe the conditional variance of Bitcoin returns. The estimation results imply that every GJR-GARCH model exhibits non-stationarity in contrast to the estimated GJR-GARCH models by Bouri et al. (2017, p. 9). This is of interest because Dyhrberg (2016) included more information in her models by including exchange rates and the stock index. The additional data could (in addition to the slightly different period under review) have led to the non-stationarity of GJR-GARCH models. The CRIX, a cryptocurrency index developed by Trimborn and Härdle (2016), was analyzed by Chen et al. (2016). The outcome of this analysis suggest an explosive ARMA(2,2)-GARCH process with student-t distributed innovations.

In the research from Ardia et al. (2019) and Radovanov et al. (2018), the Bitcoin log returns, calculated from daily average prices, are examined regarding the existence of multiple variance regimes. For this purpose, the single regime simple GARCH and GJR-GARCH models were compared with the two- and three-regime specifications of the respective models. A normal distribution, a student-t distribution, and a student-t distribution with skewness were assumed for the distribution of the model's innovations. In contrast to the following study by Caporal and Zekokh (2019), the authors clearly state that the MS-GARCH models were fitted on the residuals of an AR(1) process describing the daily Bitcoin log returns. This procedure is necessary in order to fulfil the theoretical restrictions of MS-GARCH models developed by Haas et al. (2004). The authors also concluded that the conditional variance process of Bitcoin returns exhibit structural changes.

In the research from Caporale and Zekokh (2019), which intends to extend the research of Ardia et al. (2019), a variety of MS-GARCH models are adapted to the returns of Bitcoin, Ether, Ripple, and Litecoin to determine the one-step-ahead VaR and Expected Shortfall (ES) (risk measures). The GARCH model variations considered are E-GARCH, GJR-GARCH, t-GARCH, and the simple GARCH model. For the distribution of the innovations, a standard normal distribution, a student-t distribution, and a general error distribution with and without skewness were considered. The data for historical Bitcoin prices, which the authors transformed into log returns, is obtained from the coindesk.com price index. The data set starts on 18th July 2010 and ends on 30th April 2018. In total, the authors estimated 1176 MS-GARCH models for each cryptocurrency. For the one-step-ahead prediction of VaR and ES, a rolling window approach was chosen which contained 70% of the available data. The parameters of the best MS-GARCH models were refitted (or re-estimated) daily. MS-GARCH models that did not perform well in daily back-testing (according to loss and score functions presented in the paper by Caporale and Zekokh (2019)) were not considered for further one-step-ahead predictions. The authors note that none of the single-regime GARCH models considered have succeeded in approaching the selected models. The best models consist all of two regimes. Since Caporale and Zekoh (2019) do not explicitly mention it, one cannot conclusively clarify whether more than two regimes were considered. Furthermore, the authors do not report a conditional mean equation for the log returns. However, the MS-GARCH models Caporale and Zekoh (2019) used for their empirical analysis are based on the assumption that the returns are white noise, and thus the results may be biased if this assumption is not fulfilled[v]. In any case, the authors conclude from their results that two-regime GARCH models generally provide a better predictive result for VaR and ES estimations. Furthermore, Caporale and Zekokh (2019) note that the use of single-GARCH-regime models may lead to unfavorable results for risk management, regulatory purposes, and the development of financial derivatives based on cryptocurrencies. They advocate the use of GARCH models that take multiple regimes into account (Caporale and Zekokh 2019, p. 143, 150, 154).

The last two referenced studies strongly indicate that the conditional variance of Bitcoin returns should be described using GARCH models that consider multiple regimes in order to obtain acceptable variance predictions. This may be the case today and in the near future, but not necessarily forever. Caporale et al. (2018) examined the persistence of Bitcoin, Litecoin, Ripple, and Dash returns and concluded that the cryptocurrency market is driven by inefficiency, although the inefficiency is declining (Caporale et al. 2018, p. 148). According to the Caporale et al. (2018), simple GARCH models might become more important if market efficiency improves. This could be done, for example, by integrating the cryptocurrency market into the traditional financial market to some extent. Using MS-GARCH models to describe the conditional variance of Bitcoin returns could then become obsolete. Hence, non-regime-changing GARCH models could deliver acceptable prediction results in the presence of market efficiency.

---

[v] The equation for the conditional mean of the MS-GARCH models used by Caporale and Zekokh (2019, p. 144) only contains the heteroscedastic innovation at period t. Therefore, the time series under review must be free of any autocorrelation to past events and innovations in order to obtain undistorted results (Haas et al. 2004, p. 499). Caporale and Zekokh (2019, p. 144) state that they use MS-GARCH models developed by Haas et al. (2004).

For the risk managers involved in cryptocurrencies, this means that, as long as an inefficient market can be assumed, a continuous evaluation of volatility models used to predict the conditional variance of Bitcoin returns (and other cryptocurrencies' returns) could be recommendable. Based on findings in the previous literature, this research explores whether MS-GARCH models are crucial for predicting the conditional variance of Bitcoin returns.

## 3. Methodology

### 3.1 Markov-Switching GARCH Models

The MS-GARCH models presented in this paper essentially follow Haas et al. (2004) and Ardia et al. (2016, p. 3-10). The daily log returns of Bitcoin, denoted by $r_t$, are calculated by taking the difference between the logarithmic daily closing prices $P_t$ and $P_{t-1}$.

$$r_t = \Delta \ln(P_t) = \ln(P_t) - \ln(P_{t-1})$$

Because MS-GARCH models developed by Haas et al. (2004) focus on modelling the conditional variance of a given time series, a central assumption consists of the fact that the considered variable of interest (in this case the Bitcoin log returns $r_t$) is not serially correlated, $E[r_t\, r_{t-l}] = 0$, and the expected value is zero, $E[r_t] = 0 \; \forall\, l \neq 0 \;\wedge\; t > 0$. For the time series at hand, the following equation must therefore be fulfilled[VI]:

$$r_t|(s_t = k;\ I_{t-1}) = \epsilon_{k;\,t} h_{k;\,t}^{1/2} \sim \mathcal{D}_k\big(0, h_{k;\,t}, \xi_k\big) \tag{1}$$

The conditional variance of a given regime and observation series can be generally expressed by equation (2).

$$E[r_t^2|(s_t = k;\ I_{t-1})] = h_{k;\,t} = h(r_{t-l}, h_{k;\,t-1}, \boldsymbol{\theta}_k) \tag{2}$$

---

[VI] The reader should remember that the absence of complete autocorrelation in Bitcoin log returns and possibly in many other return time series is not present, especially if the cryptocurrency market could be considered inefficient according to Caporale et al. (2018). In order to fulfil $E[r_t\, r_{t-l}] = 0$, however, the residuals of a previously fitted conditional mean equation are used instead of the log returns. Consequently, fluctuations exceeding the movements of the conditional mean are investigated with MS-GARCH models. This procedure (using a so-called de-meaned time series to analyse the excess returns of a conditional mean return series) is recommended by Ardia et al. (2016, p. 3). To be more precise, $E[r_t]$ should be expressed as $E[r_t|(s_t = k;\ I_{t-1})]$ for more accuracy in the mathematical terminology. The same applies for $E[r_t\, r_{t-l}]$. The expressions $E[r_t]$ and $E[r_t\, r_{t-l}]$ were chosen because they are more applicable for a general Introduction.

In this paper, the following GARCH models will be assumed for the function h(·).

$$h_{k;t} = \alpha_{k;0} + \alpha_{k;1} r_{t-1}^2 + \beta_k h_{k;t-1} \tag{3}$$

The E-GARCH model considers asymmetrical dependencies between past returns and the current variance. In the context of the MS-GARCH framework, the E-GARCH model is given by equation (4).

$$\ln(h_{k;t}) = \alpha_{k;0} + \alpha_{k;1}(|\epsilon_{k;t}| - E[|\epsilon_{k;t}|]) + \alpha_{k;2} r_{t-1} + \beta_k \ln(h_{k;t-1}) \tag{4}$$

The GJR-GARCH model developed by Glosten et al. (1993) is given in equation (5).

$$h_{k;t} = \alpha_{k;0} + (\alpha_{k;1} + \alpha_{k;2} I_{(r_{t-1}<0)}) r_{t-1}^2 + \beta_k h_{k;t-1} \tag{5}$$

The T-GARCH model from Zakoian (1994) focuses on the conditional volatility[VII]

$$h_{k;t}^{1/2} = \alpha_{k;0} + (\alpha_{k;1} I_{(r_{t-1}\geq 0)} - \alpha_{k;2} I_{(r_{t-1}<0)}) r_{t-1} + \beta_k h_{k;t-1}^{1/2} \tag{6}$$

Moving forward in the Markov Chain Process, the marginal density $f(r_t|\Psi, I_{t-1})$ can be expressed as a sum of the joint density functions $f_D(r_t, s_t = j| \Psi, I_{t-1})$ over all regime-switching paths (sum over the transition matrix elements). Equation (7) expresses the marginal density in terms of the probabilities and conditional density functions.

$$f(r_t|\Psi, I_{t-1}) = \sum_{i=1}^{K} \sum_{j=1}^{K} [p_{i,j} fp_{i,t-1} f_D(r_t|s_t = j, \Psi, I_{t-1})] \tag{7}$$

The value of the likelihood function L(·) is only conditioned on the set of realisations of the log returns, $\mathbf{r} = \{r_1, \ldots, r_T\}$, and depends on the model parameter vector $\Psi$. The likelihood function to be maximised is given by the equation (8). T denotes the last observation in a given log return series.

$$L(\Psi|\mathbf{r}) = \prod_{t=1}^{T} \sum_{i=1}^{K} \sum_{j=1}^{K} [p_{i,j} fp_{i,t-1} f_D(r_t|s_t = j, \Psi, I_{t-1})] \tag{8}$$

---

[VII] The terms variance and volatility are not used synonymously in context of this paper. When the term volatility is mentioned regarding a certain model, it refers to the standard deviation of a variable.

The filtered probabilities are also used as weights to obtain the overall variance or volatility. This is due to the fundamental concept of this MS-GARCH model class. The filtered probabilities are therefore the one indicator that reveals which regime is actually active. The overall conditional variance $H_t$ can therefore be approximated by $\sum_{k=1}^{K} fp_{k;t} h_{k;t}$ (Hamilton 1989, Haas et al. 2004, and Ardia et al. 2016)[VIII]. In the empirical part of this paper, all MS-GARCH model parameter vectors $\mathbf{\Psi}$ are estimated by using the previously described maximum likelihood approach.

### 3.2 Stochastic Autoregressive Volatility Models

This section follows the estimations of Fleming and Kirby (2003). The VaR estimation and prediction results in the empirical section of Fleming and Kirby's (2003. p. 394-400) paper indicate that, although SARV models can generate a better variance prediction of daily returns on exchange rates and stock indices than GARCH models, the research classifies the difference as marginal and economically meaningless. So, one might ask why this model should be used if the variance and volatility predictions are classified as marginal.

Fleming and Kirby (2003) integrate the exchange rate pairs and equity indices into the traditional financial markets, therefore the markets for these assets can be seen as efficient to a high degree with regard to the inclusion of information[IX]. But efficiency in the cryptocurrency market, by contrast, may not yet exist or might still be in development, as the previously reviewed research suggests. Therefore, SARV models might significantly outperform MS-GARCH models under such inefficient market conditions. Before outlining the SARV models, the conceptual differences between these two model classes should be briefly addressed. SARV models differ in two core aspects from GARCH and MS-GARCH models. In contrast to the GARCH model class, the function, by which the latent variable develops, contains a random component: the state innovation. The law of motions, by which the observable variable and the latent variable develops, can be expressed with simple time series models.

---

[VIII] Alternatively, the smoothed probabilities $sp_{i,t} = \Pr[s_t = i | \mathbf{\Psi}, I_T]$ can be used to obtain $H_t$. The difference between the two types of probabilities lies in the condition. The filtered probabilities in a given period $t$ are conditioned on the information set available up to $t$, $I_t$, whereas the smoothed probabilities are conditioned on the complete information set $I_T$. This indicates that future information is used in obtaining the smoothed probabilities. Since this paper intends to predict the conditional variances by using a pseudo-out-of-sample approach, future information is classified as non-available.

[IX] The authors analyzed the exchange rate of the US dollar and the following currencies: the British pound, Canadian dollar, Deutsche mark, Japanese yen, and Swiss franc. Equities were captured by analysing the NASDAQ, NYSE, S&P 500, FTSE and TOPIX indicies (Fleming and Kirby 2003, p. 395).

The observation or measurement equation in both SARV state space representations is an AR(1) process describing the conditional mean of Bitcoin log returns. Denoting them by $r_t$, equation (9) depicts this process.

$$r_t = \mu + \delta r_{t-1} + h_t^{1/2} z_t \tag{9}$$

In the first SARV model specification, the conditional variance is expressed as an AR(1) process given by equation (10).

$$h_t = \kappa + \phi h_{t-1} + \gamma h_{t-1}^{1/2} u_{t-1} \tag{10}$$

As the volatility $h_t^{1/2}$ can be obtained by taking the square root of the conditional variance $h_t$, this state space specification is known as square root SARV (SR-SARV) in the literature (Renault 2009, p. 276). The second specification is an AR(1) process for the conditional volatility given by equation (11). In contrast to the SR-SARV, the stochastic term $u_{t-1}$ in equation (11) is not multiplied by the volatility $h_{t-1}^{1/2}$.

$$h_t^{1/2} = \kappa + \phi h_{t-1}^{1/2} + \gamma u_{t-1} \tag{11}$$

For practical implementation, both SARV state space specifications can be embedded into a single Kalman filter based framework. The last steps of this algorithm predicts the variance or volatility of log returns. For this purpose, the following placeholders are used for innovations and variance or volatility.

$$y_t = \begin{cases} e_t^2 & \text{if } s_t = h_t \\ |e_t| & \text{if } s_t = h_t^{1/2} \end{cases} \tag{12}$$

Obviously, $y_t$ is the observation in the general state space framework. Considering the placeholders $y_t$ and $s_t$, both SARV models can be represented by a single general linear state space representation given by equation (13) and (14).

$$s_{t+1} = \eta + \phi(s_t - \eta) + v_t \tag{13}$$

$$y_t = a + b(s_t - \eta) + w_t \tag{14}$$

In the technical sense, the adjustment is done by using an adjustment or information weight, commonly known as the Kalman gain $K_t$. Equations (15) to (19) represent the complete Kalman algorithm (or Kalman filter) for the general state space model consisting of $y_t$ and $s_{t+1}$. The Kalman filter needs initial values to start the calculation of one-step-ahead predictions. According to Fleming and Kirby (2003, fn. 5, p. 371), the initial values $s_{1|0} = \eta$ and $P_{1|0} = Q_{1|0}(1 - \phi^2)^{-1}$ can be used.

$$s_{t|t-1} = \eta + \phi(s_{t-1|t-1} - \eta) \tag{15}$$

$$y_{t|t-1} = a + b(s_{t|t-1} - \eta) \tag{16}$$

$$P_{t|t-1} = \phi^2 \left[ P_{t-1|t-2} - b^2 P^2_{t-1|t-2}(b^2 P_{t-1|t-2} + R_{t|t-1})^{-1} \right] + Q_{t|t-1} \tag{17}$$

$$K_t = \phi b P_{t|t-1}(b^2 P_{t|t-1} + R_{t|t-1})^{-1} \tag{18}$$

$$s_{t|t} = s_{t|t-1} + \phi^{-1} K_t(y_t - y_{t|t-1}) \tag{19}$$

Once equation (19) is calculated for a given period t, there is no further obstacle in calculating the one-step-ahead prediction $s_{t+1|t}$ by using equation (15) forward lagged by one period. Equation (19) is central in the Kalman filter because it depicts the additive a-priori moment $s_{t|t-1}$ adjustment by $\phi^{-1} K_t$-times the error $y_t - y_{t|t-1}$ in order to obtain the a-posteriori moment $s_{t|t}$.

According to Fleming and Kirby (2003), the parameter constraints for both SARV models are $\kappa, \gamma \in \mathbb{R}^+, \phi \in \mathbb{R}_{\setminus\{0\}}, \mu \in \mathbb{R}$ and $\delta \in \,]-1; 1[$. Parameter estimates are obtained by numerically minimizing the target function (20) or, since the innovations of the state and observation equations are heteroscedastic and depend on the variance or volatility, minimizing the target function (20)[x]. Numerical optimization is performed by the methodology proposed by Nelder and Mead (1965). This optimization algorithm is based on a simplex and therefore offers two major advantages over gradient-based optimization algorithms. The first one is that the first derivations of the target function after the parameter vector $\boldsymbol{\theta}$ do not have to be (computationally intensively) estimated using numerical gradient estimation methods. The simplex based algorithm could therefore proceed faster in finding a (local) minima than gradient-based methods. The second advantage is the robustness in estimating (local) minima, which could deviate from the global minimum (Nelder and Mead 1965, p. 308s., 312s.). Thus, the use of multiple initial parameters is highly recommended.

---

[x] Under the assumption that no heteroscedastic innovations in the observation and state equations were present ($Q_{t|t-1} = Q \wedge R_{t|t-1} = R\ \forall t > 0$), a link was established by Fleming and Kirby (2003, p. 372s.) between GARCH models and the SARV models under review (for example, simple GARCH by Bollerslev (1986) for SR-SARV and the absolute value ARCH developed by Schwert (1990) for the AR-SARV specification). The specific variance or volatility prediction equation ($s_{t+1|t}$) could then be re-written to a GARCH type linear filter under the assumption that the Kalman gain converges to a constant ($K_t$ is time invariant). In the context of this work and in view of the extreme volatility of Bitcoin log returns, this connection is deliberately not established. The Kalman gain is assumed to be time-varying. The reason for this model design lies in the fact that the dynamics associated with heteroscedastic innovations to the state and observation equations should be sustained with the intention of extracting the latent variable as accurately as possible from the observations.

$$S_{LS} = \sum_{t=1}^{T}(y_t - y_{t|t-1})^2 \tag{20}$$

$$S_{WLS} = \sum_{t=1}^{T}(y_t - y_{t|t-1})^2 / (b^2 P_{t|t-1} + R_{t|t-1}) \tag{21}$$

Equation (20) is a least squares approach. The parameter vector $\hat{\theta}$ that minimises (20) is called the least squares (LS) estimate of the respective model parameters. For the SR-SARV model, the parameters are chosen in order to minimize the sum of $(e_t^2 - h_{t|t-1})^2$, where in the AR-SARV model the sum of $(|e_t| - \psi h_{t|t-1}^{1/2})^2$ is to be minimized by the estimator $\hat{\theta}$. However, Fleming and Kirby (2003, p. 374) note that the least squares estimates could be inefficient when both SARV models were constructed with heteroscedastic innovations in the state and observation equation. This is the case in this paper. Hence equation (21), the so-called weighted least squares (WLS) target function, will be additionally considered in the empirical section of this paper. In the WLS estimation approach, the squared prediction errors (for example, $h_{t|t-1}$ is the prediction of $e_t^2$) are normed by the conditional mean squared error $b^2 P_{t|t-1} + R_{t|t-1} = E\left[(y_t - y_{t|t-1})^2\right]$ (Fleming and Kirby 2003).

## 3.3 Data and Data Sources

The data set is obtained from the website coinmarketcap.com[XI]. The daily Bitcoin closing prices denoted in USD are transformed into daily log returns, resulting in a total of 2233 observations. The period under review begins on 29th April 2013 and extends to 9th June 2019. We use the data before the corona pandemic and the war in Ukraine due to the recent economic crises and energy crises, respectively, may present structural breaks and do not represent the normal state of the markets before the corona pandemic. Furthermore, the aim of this paper is to compare the prediction abilities of two classes of volatility models, the last 20% of the observations are

---

[XI] The data was obtained by using the web and table scraping R-package "crypto". See https://www.rdocumentation.org/packages/crypto for the documentation and further details on this package.

classified as out-of-sample data. This leaves a total of 1787 observations for the in-sample fitting of SARV and MS-GARCH models. The out-of-sample period starts on 21$^{th}$ March 2018. As a prediction approach, both an expanding and a rolling window are used where the parameters of the SARV and MS-GARCH models are re-estimated daily. In other words, after each one-step variance or volatility prediction, respectively, a new data point is available. The parameters of the best in-sample models are re-estimated in order to obtain more efficient model parameters. The rolling window inherits 800 observations. This re-estimation and prediction approach can be used when structural changes in the data-generating process (unstable parameters) are assumed to exist (Pesaran and Timmermann 2002). The next section of the study provides detailed data analysis and empirical results.

## 4. Results

In this section, we will provide a brief overview of the statistical properties of the daily log returns of the BTC/USD exchange rate. Subsequently, we will present the results for both the MS-GARCH and SARV models. Additionally, a subsection will be dedicated to discussing the unconditional mean model.

### 4.1 Explorative Data Analysis

In order to assess the distribution of Bitcoin's log returns and evaluate the suitability of ARIMA models for estimating the unconditional mean equation, we have computed descriptive statistics and conducted an Augmented Dickey-Fuller (ADF) test on the time series data for both the exchange rate and the returns. Table 2 provides an overview of the descriptive statistics for Bitcoin's log returns.

The exceptionally high value of the chi-square distributed test statistic, as proposed by Jarque and Bera (1980), results in the rejection of the null hypothesis at various commonly used significance levels. This indicates that the log returns of Bitcoin cannot be considered as following a normal distribution. Furthermore, the distribution of returns displays heavy-tailed characteristics.

Table 2: Descriptive statistics for the Bitcoin log returns of the entire sample.

| Statistic | Value |
|---|---|
| Mean | 0.0018 |
| Median | 0.0019 |
| Standard deviation | 0.0431 |
| Skewness | - 0.1543 |
| Kurtosis | 10.9270 |
| Maximum log. return | 0.3574 |
| Minimum log. return | - 0.2662 |
| Jarque-Bera test statistic | 5855.13 |

Table 3 presents the results of the Augmented Dickey-Fuller (ADF) tests for both the USD/BTC exchange rate series and the Bitcoin log returns. Unlike the exchange rate time series, the returns are found to be stationary at the 1% significance level.

Table 3: Summary of the augmented Dickey-Fuller test for the exchange rate and log returns

| | Exchange rate (USD/BTC) | BTC returns |
|---|---|---|
| ADF test statistic | 0.40 | - 8.41 |
| ADF critical value 5% | - 3.41 | - 2.86 |
| ADF critical value 1% | - 3.96 | - 3.43 |
| Included lags | 90 | 16 |

### 4.2 Results for the Conditional Mean

Table 4 reports the best ARMA models, corresponding information criteria and additionally the Ljung-Box test statistic for the residuals of the respective ARMA model.

Table 4: Results for the Conditional Mean

| Constant | AR | MA | AIC | BIC | HQ | Ljung-Box test statistic | Ljung-Box critical value 5% | Ljung-Box critical value 1% |
|---|---|---|---|---|---|---|---|---|
| TRUE | 6 | 5 | **- 3.3741** | - 3.3372 | - 1.6803 | 34.75 | 28.87 | 34.81 |
| FALSE | 5 | 6 | - 3.3740 | **- 3.3402** | **- 1.6808** | 35.04 | 30.14 | 36.19 |
| FALSE | 6 | 5 | - 3.3739 | - 3.3401 | -1.6807 | 35.36 | 30.14 | 36.19 |
| TRUE | 7 | 5 | - 3.3735 | - 3.3336 | - 1.6794 | 33.95 | 27.59 | 33.41 |
| TRUE | 5 | 7 | - 3.3735 | - 3.3335 | - 1.6794 | 33.97 | 27.59 | 33.41 |
| TRUE | 5 | 8 | - 3.3725 | - 3.3295 | - 1.6783 | 34.18 | 26.30 | 31.99 |
| TRUE | 8 | 5 | - 3.3725 | - 3.3295 | - 1.6783 | 34.02 | 26.30 | 31.99 |
| TRUE | 5 | 9 | - 3.3723 | - 3.3262 | - 1.6777 | 31.85 | 24.99 | 30.58 |
| TRUE | 9 | 5 | - 3.3722 | - 3.3261 | - 1.6776 | 32.01 | 24.99 | 30.58 |
| FALSE | 9 | 5 | - 3.3719 | - 3.3289 | - 1.6780 | 32.75 | 26.30 | 31.99 |

In the column Constant, TRUE means that a constant has been taken into account in the ARMA model. AR and MA indicate the order of the ARMA model. AIC (BIC, HQ) is the information criterion of Akaike (Bayes-Schwarz, Hannan-Quinn).

The estimated model parameters are reported in table 5. The standard errors are in parentheses[XII].

Table 5: Estimated Model Parameters

ARMA (5, 6)

| Order | AR | MA |
|---|---|---|
| Constant | - | - |
| 1 | - 0.0749 (0.0594) | 0.0800 (0.0635) |
| 2 | 0.3982*** (0.0683) | - 0.4143*** (0.0666) |
| 3 | - 0.4927*** (0.0435) | 0.4887*** (0.0454) |
| 4 | 0.1114** (0.0470) | - 0.0641 (0.0522) |
| 5 | 0.8377*** (0.0510) | - 0.8166*** (0.0514) |
| 6 | - | 0.0506* (0.0292) |

ARMA (6, 5)

| Order | AR | MA |
|---|---|---|
| Constant | 0.0023 (0.0014) | |
| 1 | - 0.1437** (0.0668) | 0.1491** (0.0632) |
| 2 | 0.3815*** (0.0735) | - 0.3964*** (0.0729) |
| 3 | - 0.4756*** (0.0499) | 0.4694*** (0.0476) |
| 4 | 0.0696 (0.0550) | - 0.0247 (0.0567) |
| 5 | 0.8340*** (0.0531) | - 0.8106*** (0.0548) |
| 6 | 0.0494* (0.0290) | - |

Significance levels are denoted by */ **/ ***, implying significance at the 10%, 5% and 1 % level respectively

---

[XII] The vector of the standard errors $\mathbf{SE} = [se_i]$ was calculated on the basis of the Hessian matrix $\mathbf{H}$ of the parameters of the respective model estimation. In this case, the inverse of the Hessian matrix $\mathbf{H^{-1}}$ is the m x m dimentional covariance matrix of the m parameters. The square roots of the diagonal elements of the covariance matrix are then the estimated standard errors of the model parameters. On the basis of these SE, the t-values can be calculated to check the statistical significance of a parameter. The standard error of the i-th element of $\mathbf{SE}$ is calculated according to $se_i = [cvar_{i,i}]^{1/2}$, where $cvar_{i,i}$ is the i-th diagonal element of $\mathbf{H^{-1}}$. $i \in \{1, ..., m\}$. */ **/ *** imply significance at the 10%, 5% and 1 % level respectively.

The comparison reveals that the AR and MA parameters, significant at the 1% level, have almost the same size in both ARMA models. Therefore, these models share similar characteristics and can be considered stationary. Stationarity of ARMA models is solely based on the AR parameters sum (see Tsay (2010) for example). In both models, the AR parameters sums (0.7797 for ARMA(5,6) and 0.7152 for ARMA(6,5)) are less than one. Finally, based on the ARMA models' residuals, MS-GARCH models were fitted and presented in the next section of results.

**4.3 Results for the Estimation of the Markov-Switching GARCH Models**

Table 6 lists the best MS-GARCH model combinations fitted to the residuals of the AIC optimal ARMA(6,5) model.

Table 6: MS-GARCH model combinations fitted to the residuals of the AIC optimal ARMA(6,5) model.

| GARCH regime one | GARCH regime two | Innovation distribution regime one | Innovation distribution regime two | AIC | BIC | HQ | Ljung-Box test statistic | Ljung-Box critical value 5% | Ljung-Box critical value 1% |
|---|---|---|---|---|---|---|---|---|---|
| sGARCH | sGARCH | sstd | sstd | **-4.0147** | -3.9779 | **-2.0005** | 37.23 | 43.77 | 50.89 |
| eGARCH | sGARCH | sstd | sstd | -4.0145 | -3.9746 | -1.9999 | 36.21 | 44.98 | 52.19 |
| eGARCH | gjrGARCH | sstd | sstd | -4.0140 | -3.9710 | -1.9990 | 36.13 | 46.19 | 53.48 |
| sGARCH | gjrGARCH | sstd | sstd | -4.0137 | -3.9738 | -1.9994 | 37.26 | 44.98 | 52.19 |
| eGARCH | sGARCH | std | sstd | -4.0133 | -3.9764 | -1.9998 | 35.61 | 43.77 | 50.89 |
| eGARCH | gjrGARCH | std | sstd | -4.0132 | -3.9733 | -1.9992 | 36.45 | 44.98 | 52.19 |
| eGARCH | sGARCH | std | snorm | -4.0128 | -3.9790 | -2.0002 | 33.82 | 42.55 | 49.58 |
| eGARCH | sGARCH | sstd | snorm | -4.0120 | -3.9752 | -1.9992 | 33.55 | 43.77 | 50.89 |
| eGARCH | gjrGARCH | std | snorm | -4.0119 | -3.9750 | -1.9991 | 33.81 | 43.77 | 50.89 |
| sGARCH | eGARCH | std | sstd | -4.0116 | -3.9748 | -1.9990 | 35.55 | 43.77 | 50.89 |
| sGARCH | eGARCH | sstd | sstd | -4.0115 | -3.9716 | -1.9983 | 35.36 | 44.98 | 52.19 |
| eGARCH | gjrGARCH | sstd | snorm | -4.0110 | -3.9711 | -1.9981 | 33.74 | 44.98 | 52.19 |
| eGARCH | eGARCH | sstd | sstd | -4.0109 | -3.9679 | -1.9975 | 34.15 | 46.19 | 53.48 |
| eGARCH | eGARCH | std | snorm | -4.0108 | -3.9740 | -1.9986 | 34.02 | 43.77 | 50.89 |
| sGARCH | sGARCH | std | snorm | -4.0106 | **-3.9799** | -1.9996 | 35.87 | 41.33 | 48.27 |

Observing the individual GARCH specifications in the respective regimes, one notices that the T-GARCH model is not included under the top ARMA(6,5) MS-GARCH models. Instead, the E-GARCH and GJR-GRACH models are present. This could be an indication that the asymmetry modelling and the limitations of the parameters in the T-GARCH model do not suitably describe the conditional volatility of the ARMA residuals, which can also be called excess returns. Remarkably, all information criteria prefer the simple GARCH model. The distribution of standardized innovations, however, is mixed. The AIC and HQ criteria prefer a student-t distribution with skewness in both regimes. The BIC criterion, on the other hand, recommends a student-t distribution in the first regime and a normal distribution with skewness in the second. So, according to the AIC and HQ criteria, skewness and heavy tails prevail in both volatility regimes, while according to BIC, skewness exists only in the second regime and in the first heavy tails. The values of the Ljung-Box test statistics are lower than the critical values at the 5 % and 1 % levels for all reported models in table 6. Therefore, by accepting the Ljung-Box null hypothesis, one can conclude that the residuals of all reported models are free of serial correlations.

Table 7 illustrates the model parameters and standard errors of the best ARMA(6,5) MS-GARCH models.

Table 7: Parameters and Corresponding Standard Errors

|  | ARMA(6,5) MS-sGARCH-sstd-sstd | | | ARMA(6,5) MS-sGARCH-std-snorm | |
|---|---|---|---|---|---|
|  | k = 1 | k = 2 |  | k = 1 | k = 2 |
|  | sGARCH | sGARCH |  | sGARCH | sGARCH |
| $\alpha_{k;0}$ | 0.0000*** | 0.0001*** | $\alpha_{k;0}$ | 0.0000*** | 0.0002*** |
|  | (0.0000) | (0.0000) |  | (0.0000) | (0.0000) |
| $\alpha_{k;1}$ | 0.0664*** | 0.1740*** | $\alpha_{k;1}$ | 0.1105*** | 0.1035*** |
|  | (0.0032) | (0.0048) |  | (0.0024) | (0.0029) |
| $\beta_k$ | 0.9306*** | 0.8173*** | $\beta_k$ | 0.8884*** | 0.8627*** |
|  | (0.0001) | (0.0003) |  | (0.0000) | (0.0009) |
| $\xi_k$ | 1.0419*** | 0.8499*** | $\xi_k$ | - | 0.7222*** |
|  | (0.0008) | (0.0010) |  | - | (0.0015) |
| $\nu_k$ | 2.4545*** | 4.5263*** | $\nu_k$ | 2.6915*** | - |
|  | (0.0026) | (0.0199) |  | (0.0029) | - |
| $P^T$ | 0.9679*** | 0.0321 | $P^T$ | 0.9716*** | 0.0284 |
|  | (0.0003) |  |  | (0.0007) |  |
|  | 0.0344*** | 0.9659 |  | 0.0799*** | 0.9201 |
|  | (0.0003) |  |  | (0.0003) |  |

Significance levels are denoted by */ **/ ***, implying significance at the 10%, 5% and 1 % level respectively

The parameter vector $\Psi$ and the corresponding standard errors are presented in table 7[XIII] for both selections, best BIC as well as best AIC and HQ ARMA MS-GARCH models. Afterwards, the best MS-GARCH models are fitted on the residuals of the ARMA(5,6) process.

As a preliminary conclusion of the MS-GARCH model estimation, one can state that the optimal volatility models adapted to the residuals of the ARMA(6,5) process are partially very

---

[XIII] The extremely low standard error and consequently high significance of all model parameters could be unusual. This may be related to the methodology used by the MSGARCH R package to calculate the standard errors. If the standard errors are calculated according to the conventional method (see fn. 35 of this paper), no parameter is statistically significant. Since statistical significance is important for the in-sample fit to be able to identify relationships, this does not necessarily have to be important for the out-of-sample variance prediction. Hence, this conspicuousness is not discussed in more detail in this paper.

similar to each other. The transition probabilities $p_{11}$ and $p_{22}$ as well as $H_t$, $h_{1,t}$, and $h_{2,t}$ are highly persistent in the AIC and HQ, and BIC optimal model constellations. In the BIC optimal model, a distinction is made between a skewed normal distribution for $k = 2$ and a student-t distribution for $k = 1$, so the ARMA(6,5) MS-GARCH model requires two fewer parameters than the AIC- and HQ-preferred MS-GARCH model, namely one skewness parameter in $k = 1$ and one shape parameter in $k = 2$. However, the course of $H_t$ in the BIC optimal model is slightly dampened compared to the course of $H_t$ in the AIC and HQ optimal model. For this reason, all optimal ARMA MS-GARCH model constellations are compared in the prediction section of this paper.

This subchapter concludes by summarizing the discussion on the optimal ARMA(5,6) MS-GARCH models. Section 4.2 already concluded that the AIC-preferred ARMA(6,5) and the BIC- and HQ-preferred ARMA(5,6) models for describing the conditional mean of Bitcoin log returns are very similar, so clearly committing to one conditional mean model for the MS-GARCH section was not possible. Accordingly, the result for the MS-GARCH models fitted on the residuals of the ARMA(5,6) model are analogous to the observations already reported for the ARMA(6,5) MS-GARCH model constellation. For the sake of completeness, all figures and tables for the ARMA(5,6) MS-GARCH model are reported below. The conclusions drawn from tables 8 and 9 are analogous to those from tables 6 and 7. The AIC and HQ information criteria imply a simple GARCH model in both regimes. A student-t distribution with skewness is identified as the best fit for $\mathcal{D}_k$ in both regimes. The BIC, on the other hand, prefers different distributions for the standardized innovations $\epsilon_{1;t}$ and $\epsilon_{2;t}$, namely a skewed standard normal distribution in $k = 2$ (snorm) and a symmetric heavy tailed distribution in $k = 1$ (std). Table 9 additionally reveals that all reported model parameters show almost no differences to those in table 7. The observation made based on figure 4 panel C and figure 5 panel C (i.e., that in the BIC-preferred ARMA(6,5) MS-GARCH model the course of $H_t$ in relation to the AIC- and BIC-preferred model appears to be dampened) can also be adopted for the ARMA(5,6) MS-GARCH model. See figure 7 panel C and figure 8 panel C for details. Also, the dominance of the first regime according to the Viterbi algorithm in the BIC-preferred ARMA(5,6) MS-GARCH model can be determined by examining figure 8 in relation to figure 5. Furthermore, with regard to the model diagnosis of the ARMA(5,6) MS-GARCH models, there is also no serial correlation in the ARMA(6,5) MS-GARCH models' residuals.

Table 8 provides the best ARMA(5,6), MS-GARCH models, corresponding information criteria and Ljung-Box test statistics for the model's residuals.

Table 8: MS-GARCH model combinations that have been fitted to the residuals of the BIC and HQ optimal ARMA(5,6) model.

| GARCH regime one | GARCH regime two | Innovation distribution regime one | Innovation distribution regime two | AIC | BIC | HQ | Ljung-Box test statistic | Ljung-Box critical value 5% | Ljung-Box critical value 1% |
|---|---|---|---|---|---|---|---|---|---|
| sGARCH | sGARCH | sstd | sstd | **- 4.0140** | - 3.9771 | **- 2.0002** | 36.43 | 44.98 | 52.19 |
| sGARCH | gjrGARCH | sstd | sstd | - 4.0131 | - 3.9731 | - 1.9991 | 36.60 | 46.19 | 53.48 |
| eGARCH | sGARCH | std | sstd | - 4.0119 | - 3.9750 | - 1.9991 | 33.46 | 44.98 | 52.19 |
| sGARCH | eGARCH | std | sstd | - 4.0118 | - 3.9750 | - 1.9991 | 35.03 | 44.98 | 52.19 |
| eGARCH | sGARCH | sstd | sstd | - 4.0114 | - 3.9714 | - 1.9983 | 32.95 | 46.19 | 53.48 |
| sGARCH | eGARCH | sstd | sstd | - 4.0112 | - 3.9713 | - 1.9982 | 35.05 | 46.19 | 53.48 |
| eGARCH | gjrGARCH | std | sstd | - 4.0109 | - 3.9710 | - 1.9980 | 33.55 | 46.19 | 53.48 |
| eGARCH | sGARCH | sstd | snorm | - 4.0106 | - 3.9737 | - 1.9985 | 34.21 | 44.98 | 52.19 |
| eGARCH | gjrGARCH | sstd | sstd | - 4.0104 | - 3.9674 | - 1.9972 | 33.10 | 47.39 | 54.77 |
| sGARCH | tGARCH | std | sstd | - 4.0101 | - 3.9732 | - 1.9982 | 34.00 | 44.98 | 52.19 |
| eGARCH | sGARCH | std | snorm | - 4.0100 | - 3.9762 | - 1.9987 | 33.85 | 43.77 | 50.89 |
| eGARCH | gjrGARCH | sstd | snorm | - 4.0097 | - 3.9697 | - 1.9974 | 34.03 | 46.19 | 53.48 |
| tGARCH | sGARCH | std | sstd | - 4.0094 | - 3.9726 | - 1.9979 | 36.92 | 44.98 | 52.19 |
| eGARCH | eGARCH | std | sstd | - 4.0092 | - 3.9693 | - 1.9972 | 32.61 | 46.19 | 53.48 |
| eGARCH | gjrGARCH | std | snorm | - 4.0091 | - 3.9722 | - 1.9977 | 33.72 | 44.98 | 52.19 |
| eGARCH | eGARCH | sstd | snorm | - 4.0088 | - 3.9689 | - 1.9970 | 34.29 | 46.19 | 53.48 |
| sGARCH | sGARCH | sstd | snorm | - 4.0086 | - 3.9749 | - 1.9981 | 35.55 | 43.77 | 50.89 |
| tGARCH | eGARCH | snorm | sstd | - 4.0086 | - 3.9687 | - 1.9969 | 34.64 | 46.19 | 53.48 |
| eGARCH | eGARCH | sstd | sstd | - 4.0086 | - 3.9656 | - 1.9963 | 32.65 | 47.39 | 54.77 |
| tGARCH | gjrGARCH | std | sstd | - 4.0085 | - 3.9686 | - 1.9969 | 36.98 | 46.19 | 53.48 |
| sGARCH | sGARCH | std | snorm | - 4.0084 | **- 3.9777** | - 1.9985 | 35.59 | 42.55 | 49.58 |

The table 9 describes the parameter vector $\Psi$ and corresponding errors of the best models according AIC, BIC, HQ criterion.

Table 9: The parameter vector $\Psi$ and corresponding standard errors of the best models according to AIC, BIC, and HQ.

| | ARMA(5,6) MS-sGARCH-sstd-sstd | | | ARMA(5,6) MS-sGARCH-std-snorm | |
|---|---|---|---|---|---|
| | k = 1 | k = 2 | | k = 1 | k = 2 |
| | sGARCH | sGARCH | | sGARCH | sGARCH |
| $\alpha_{k;0}$ | 0.0000*** (0.0000) | 0.0002*** (0.0000) | $\alpha_{k;0}$ | 0.0000*** (0.0000) | 0.0002*** (0.0000) |
| $\alpha_{k;1}$ | 0.0646*** (0.0039) | 0.1738*** (0.0042) | $\alpha_{k;1}$ | 0.1122*** (0.0025) | 0.1169*** (0.0034) |
| $\beta_k$ | 0.9341*** (0.0001) | 0.8185*** (0.0002) | $\beta_k$ | 0.8867*** (0.0000) | 0.8565*** (0.0008) |
| $\xi_k$ | 0.9692*** (0.0007) | 0.8314*** (0.0010) | $\xi_k$ | - - | 0.7102*** (0.0014) |
| $\nu_k$ | 2.4577*** (0.0024) | 4.4294*** (0.0186) | $\nu_k$ | 2.6775*** (0.0027) | - - |
| $\mathbf{P^T}$ | 0.9637*** (0.0003) 0.0376*** (0.0003) | 0.0363 0.9624 | $\mathbf{P^T}$ | 0.9694*** (0.0007) 0.0790*** (0.0003) | 0.0306 0.9210 |

Significance levels are denoted by */ **/ ***, implying significance at the 10%, 5% and 1 % level respectively

## 4.4 Estimation of Stochastic Autoregressive Volatility Models

Table 10 reports the best SR-SARV models, initial values, and least square estimators $\hat{\theta}$ for these. The values were obtained by minimizing $S_{LS}$.

Table 10: Results on the best SR-SARV models

| Initial values for optimisation | | | | | | Least square estimator | | | | |
|---|---|---|---|---|---|---|---|---|---|---|
| κ | φ | γ | μ | δ | Target function (LS) | $\hat{\kappa}$ | $\hat{\phi}$ | $\hat{\gamma}$ | $\hat{\mu}$ | $\hat{\delta}$ |
| 0.4 | 0.4 | 0.7 | -0.2 | 0.5 | 0.06354 | 0.00076 | 0.62454 | 1.22412 | 0.00002 | 0.09817 |
| | | | | | | (0.02134) | (8.77993) | (3.34328) | (0.17593) | (1.85438) |
| 0.1 | 0.7 | 0.7 | 0.5 | 0.5 | 0.06356 | 0.00062 | 0.69304 | 0.98839 | -0.00012 | 0.09502 |
| | | | | | | (0.01866) | (7.59049) | (23.4589) | (0.17768) | (1.87008) |
| 0.1 | 0.7 | 0.7 | 0.5 | -0.2 | 0.06357 | 0.00062 | 0.69553 | 0.97077 | -0.00009 | 0.09400 |
| | | | | | | (0.01843) | (7.49546) | (22.6037) | (0.17786) | (1.86820) |
| 0.4 | 0.7 | 0.7 | 0.5 | -0.2 | 0.06357 | 0.00089 | 0.57420 | 1.40326 | 0.00009 | 0.10174 |
| 0.1 | 0.4 | 0.7 | -0.9 | -0.2 | 0.06357 | 0.00086 | 0.58131 | 1.41010 | 0.00010 | 0.10772 |
| 0.1 | 0.4 | 0.7 | -0.2 | 0.5 | 0.06357 | 0.00061 | 0.69695 | 0.95834 | -0.00015 | 0.09298 |
| 0.1 | 0.1 | 0.4 | 0.5 | 0.5 | 0.06358 | 0.00087 | 0.56992 | 1.46642 | 0.00021 | 0.10451 |
| 0.1 | 0.1 | 0.7 | 0.5 | -0.2 | 0.06358 | 0.00060 | 0.70672 | 0.93019 | -0.00023 | 0.09284 |
| 0.1 | 0.4 | 0.7 | 0.5 | 0.5 | 0.06359 | 0.00068 | 0.66316 | 0.99655 | -0.00014 | 0.09119 |
| 0.4 | 0.1 | 0.7 | 0.5 | 0.5 | 0.06359 | 0.00091 | 0.55801 | 1.56996 | 0.00034 | 0.10397 |

According to the table 10 the least square estimates of the SR-SARV models result in a stable value of the target function $S_{LS}$ for all reported models. The values of $S_{LS}$ vary only from the fifth decimal place onward. If one considers the least square estimator $\hat{\theta}$, a slightly different picture emerges. The estimated parameters vary in dependence of the initial values. The target function $S_{LS}$ thus features multiple minima. This could be a problem regarding the predicting abilities of the estimated SARV models. The standard errors (in parentheses) indicate that no estimated parameter is significant at the levels commonly used in the literature (10%, 5% and 1% error probability). This observation strongly contrasts with the MS-GARCH models, where all estimated parameters are significant at the 0.1% level.

A different picture emerges in table 11 with regard to the target function $S_{WLS}$.

Table 11: Estimation of best SR-SARV models with weighted least squares

| Initial values for optimisation | | | | | | Weighted least square estimator | | | | |
|---|---|---|---|---|---|---|---|---|---|---|
| κ | φ | γ | μ | δ | Target function (WLS) | $\hat{\kappa}$ | $\hat{\phi}$ | $\hat{\gamma}$ | $\hat{\mu}$ | $\hat{\delta}$ |
| 0.1 | 0.7 | 0.4 | -0.9 | -0.9 | 0.00005 | 44.6185 | 0.99070 | 696.63 | -69.6466 | -0.09815 |
| | | | | | | (8879.68) | (NaN) | (43775) | (NaN) | (104.04) |
| 0.7 | 0.1 | 0.1 | -0.9 | 0.5 | 0.00056 | 0.00367 | 1.6574 | 170.92 | 0.03110 | -0.3925 |
| | | | | | | (0.2721) | (186.85) | (3165.4) | (5.5106) | (37.604) |
| 0.4 | 0.4 | 0.1 | -0.9 | -0.9 | 0.13238 | 0.99631 | 0.37191 | 124.78 | 0.59801 | 12.4346 |
| | | | | | | (NaN) | (2.5148) | (NaN) | (3.0099) | (NaN) |
| 0.4 | 0.7 | 0.7 | 0.5 | -0.2 | 1.97903 | 0.00420 | 0.80560 | 1.8615 | -0.07422 | -0.56244 |
| 0.1 | 0.4 | 0.4 | 0.5 | 0.5 | 5.07543 | 0.03441 | 0.91558 | 0.8280 | 0.61649 | -0.35932 |

Table 12 reports the results for the AR-SARV estimators. The target function is $S_{LS}$. The AR-SARV estimators systematically produce higher target function values relative to the reported SR-SARV estimators, which suggests that these model do not fit the Bitcoin log returns as well as the SR specification. The standard errors could not be calculated for all parameter estimates (denoted by NaN).

Table 12 Results of AR-SARV Estimators

| Initial values for optimisation | | | | | | Least square estimator | | | | |
|---|---|---|---|---|---|---|---|---|---|---|
| κ | φ | γ | μ | δ | Target function (LS) | $\hat{\kappa}$ | $\hat{\phi}$ | $\hat{\gamma}$ | $\hat{\mu}$ | $\hat{\delta}$ |
| 0.4 | 0.5 | 0.1 | -0.2 | -0.2 | 1.9293 | 0.01936 | 0.53815 | 0.13684 | -0.01388 | -0.04295 |
| | | | | | | (NaN) | (NaN) | (NaN) | (0.02512) | (0.35906) |
| 0.7 | 0.5 | 0.1 | -0.2 | 0.5 | 1.9337 | 0.01988 | 0.53391 | 0.15899 | -0.01451 | -0.05131 |
| | | | | | | (NaN) | (NaN) | (NaN) | (0.04362) | (0.42155) |
| 0.1 | -0.2 | 0.7 | 0.5 | -0.9 | 1.9391 | 0.00651 | 0.86309 | 0.05066 | -0.01991 | -0.20343 |
| | | | | | | (NaN) | (NaN) | (NaN) | (0.02081) | (0.37290) |
| 0.4 | 0.5 | 0.7 | -0.2 | -0.2 | 1.9489 | 0.02279 | 0.40823 | 0.21666 | -0.00805 | 0.06720 |
| 0.7 | 0.5 | 0.1 | -0.2 | -0.9 | 1.9502 | 0.02060 | 0.43672 | 0.36778 | -0.00024 | 0.17846 |
| 0.4 | 0.5 | 0.4 | -0.2 | 0.5 | 1.9561 | 0.02240 | 0.38724 | 0.44262 | -0.00309 | 0.10903 |
| 0.1 | -0.2 | 0.4 | 0.5 | 0.5 | 1.9562 | 0.02200 | 0.39144 | 0.46560 | -0.00003 | 0.12700 |
| 0.1 | 0.5 | 0.7 | 0.5 | -0.9 | 1.9573 | 0.02640 | 0.36041 | 0.33856 | -0.01321 | -0.00388 |
| 0.1 | 0.5 | 0.1 | -0.9 | -0.9 | 1.9575 | 0.02449 | 0.37306 | 0.31072 | -0.00970 | 0.03847 |
| 0.4 | 0.5 | 0.4 | -0.2 | -0.2 | 1.9602 | 0.02687 | 0.40645 | 0.18474 | -0.01880 | -0.09362 |

The table 13 lists the parameter estimates for the AR-SARV specification. The target function is $S_{WLS}$. The information contained in table 13 was discussed in relation to table 11. The focus lies on the estimates $\hat{\mu}$ and $\hat{\gamma}$.

Table 13: Estimation of parameter for the AR-SARV specification.

| Initial values for optimisation | | | | | | Weighted least square estimator | | | | |
|---|---|---|---|---|---|---|---|---|---|---|
| κ | φ | γ | μ | δ | Target function (WLS) | $\hat{\kappa}$ | $\hat{\phi}$ | $\hat{\gamma}$ | $\hat{\mu}$ | $\hat{\delta}$ |
| 0.4 | 0.5 | 0.7 | -0.9 | -0.2 | 0.000000 | 5246.3 | 0.47988 | 3462.94 | -8048.40 | -0.37037 |
| | | | | | | (NA) | (NA) | (NA) | (NA) | (NA) |
| 0.4 | 0.5 | 0.7 | -0.9 | -0.9 | 0.000003 | 67.5695 | 0.87585 | 2195.67 | -434.202 | 0.56446 |
| | | | | | | (NaN) | (0.00006) | (984294) | (NaN) | (1054.61) |
| 0.4 | 0.5 | 0.4 | 0.5 | -0.2 | 0.000004 | 169.609 | 0.78238 | 1355.86 | 621.808 | -0.28169 |
| | | | | | | (NaN) | (NaN) | (290850) | (NaN) | (327.355) |
| 0.4 | 0.5 | 0.4 | -0.9 | 0.5 | 0.000005 | 558.461 | 0.62989 | 1532.88 | -1204.01 | 0.20713 |
| 0.4 | 0.5 | 0.1 | -0.9 | -0.9 | 0.000006 | 116.054 | 0.94027 | 1327.44 | -1550.21 | 0.01983 |
| 0.4 | 0.5 | 0.1 | 0.5 | -0.9 | 0.000015 | 197.996 | 0.35697 | 700.167 | 245.712 | 0.11088 |
| 0.4 | 0.5 | 0.1 | -0.9 | 0.5 | 0.000016 | 438.384 | 0.46188 | 644.005 | -649.981 | -0.37740 |
| 0.4 | 0.5 | 0.7 | -0.9 | 0.5 | 0.000020 | 152.106 | 0.65610 | 704.922 | -352.929 | 0.20784 |
| 0.4 | -0.2 | 0.4 | -0.9 | -0.2 | 0.000038 | 2.25768 | 0.86021 | 581.252 | -13.1442 | -0.89311 |
| 0.4 | 0.5 | 0.7 | 0.5 | -0.2 | 0.000045 | 1787.98 | 0.78411 | 10190.6 | 6607.93 | 22.184 |

Further, in order to evaluate the quality of the variance predictions between the two model classes, three loss functions are evaluated for each best MS-GARCH and SARV model. The choice of the loss functions examined in this paper is based on the usual error measures in the literature for volatility predictions, e.g., Brownlees et al. (2011). The loss functions being used in this paper are the mean absolute error (MAE), the mean squared error (MSE), and the quasi likelihood (QL) given by equations (22), (23), and (24).

$$\text{MAE} = N^{-1} \sum_{t=1}^{N} \left| \hat{\sigma}_t^2 - h_{t|t-k} \right| \tag{22}$$

$$\text{MSE} = N^{-1} \sum_{t=1}^{N} \left( \hat{\sigma}_t^2 - h_{t|t-k} \right)^2 \tag{23}$$

$$\text{QL} = N^{-1} \sum_{t=1}^{N} \left( \hat{\sigma}_t^2 / h_{t|t-k} - \ln \hat{\sigma}_t^2 / h_{t|t-k} - 1 \right) \tag{24}$$

The conditional variance of Bitcoin log returns is an unobservable variable. For this reason, an approximation of the conditional variance is required in order to be able to compare the model's predictions. In equations (22) to (24), $\hat{\sigma}_t^2$ is an ex-post approximation of the true conditional variance. Brownlees et al. (2011) recommend that researchers use the squared log returns $r_t^2$ for it. N denotes the number of out-of-sample observations and $h_{t|t-k}$ is the respective model's (SARV or MS-GARCH) conditional variance prediction based on the set of information available at the time $t - k$. In this paper only one-step-ahead predictions of the conditional variance or volatility are calculated so that k equals one.

Table 14 reports the variance prediction errors of these models for the expanding window approach.

Table 14 Variance prediction error of the models.

| Forecast model | | | | Forecast errors | | |
|---|---|---|---|---|---|---|
| Conditional mean | Conditional variances | Innovation distributions | Conditional variance type | MAE | MSE | QL |
| ARMA(6,5) | sGARCH | sstd | Regime one | 0.00140 | 0.0000075 | 2.3975 |
| | sGARCH | sstd | Regime two | 0.00187 | 0.0000081 | 2.2209 |
| | | | Overall variance | 0.00163 | 0.0000078 | 3.1982 |
| ARMA(6,5) | sGARCH | std | Regime one | 0.00143 | 0.0000075 | 2.4572 |
| | sGARCH | snorm | Regime two | 0.00232 | 0.0000091 | 2.3627 |
| | | | Overall variance | 0.00157 | 0.0000077 | 3.2555 |
| ARMA(5,6) | sGARCH | sstd | Regime one | 0.00139 | 0.0000075 | 2.4476 |
| | sGARCH | sstd | Regime two | 0.00192 | 0.0000082 | 2.2313 |
| | | | Overall variance | 0.00160 | 0.0000077 | 3.3374 |
| ARMA(5,6) | sGARCH | std | Regime one | 0.00143 | 0.0000075 | 2.4644 |
| | sGARCH | snorm | Regime two | 0.00223 | 0.0000089 | 2.3336 |
| | | | Overall variance | 0.00157 | 0.0000077 | 3.2668 |
| | SR - SARV | - | - | 0.00170 | 0.0000082 | 2.1844 |
| | AR - SARV | - | - | 0.00166 | 0.0000078 | 2.1552 |

According to the QL, the overall variance prediction of the MS-GARCH models has higher errors than the predictions of the particular volatility regimes. According to the QL loss, the second regime of the ARMA(6,5) MS-GARCH model is best suited to predict the conditional variance of Bitcoin log returns.

Based on the MAE and the MSE losses, the SARV models differ only marginally from MS-GARCH models. A different conclusion can be drawn by using the QL loss as a selection criterion. A QL loss of 2.18 and 2.15 for the SR-SARV and AR-SARV (respectively) indicates the superiority of SARV models over all MS-GARCH models. Although the differences between the SARV models are relatively small, the AR-SARV specification is preferred over the SR-SARV for each loss function.

Table 15 lists the prediction errors of MS-GARCH and SARV models for the rolling window approach.

Table 15: Estimation of the prediction errors of MS-GARCH and SARV models

| Forecast model | | | | Forecast errors | | |
|---|---|---|---|---|---|---|
| Conditional mean | Conditional variances | Innovation distributions | Conditional variance type | MAE | MSE | QL |
| ARMA(6,5) | sGARCH | sstd | Regime one | 0.00143 | 0.0000075 | 2.2350 |
| | sGARCH | sstd | Regime two | 0.00196 | 0.0000082 | 2.2706 |
| | | | Overall variance | 0.00167 | 0.0000077 | 2.9704 |
| ARMA(6,5) | sGARCH | std | Regime one | 0.00142 | 0.0000075 | 2.2524 |
| | sGARCH | snorm | Regime two | 0.00219 | 0.0000086 | 2.3369 |
| | | | Overall variance | 0.00164 | 0.0000076 | 2.9670 |
| ARMA(5,6) | sGARCH | sstd | Regime one | 0.00146 | 0.0000075 | 2.2451 |
| | sGARCH | sstd | Regime two | 0.00194 | 0.0000082 | 2.2736 |
| | | | Overall variance | 0.00167 | 0.0000077 | 2.9973 |
| ARMA(5,6) | sGARCH | std | Regime one | 0.00141 | 0.0000075 | 2.2804 |
| | sGARCH | snorm | Regime two | 0.00215 | 0.0000086 | 2.3304 |
| | | | Overall variance | 0.00162 | 0.0000076 | 3.0150 |
| | SR - SARV | - | - | 0.00180 | 0.0000085 | 2.2681 |
| | AR - SARV | - | - | 0.06649 | 0.0091021 | 4.1458 |

Excluding past information, as it is the case in the rolling window approach, improves the variance predictions of MS-GARCH models and worsens the predictions of the SARV models. The QL losses for all reported MS-GARCH models decreased compared to the metrics reported in table 14. Under the rolling window, the best performing MS-GARCH model is the one with the lowest BIC (ARMA(6,5) MS-GARCH model). For this model, the QL loss of the overall variance is 2.9670.

Under the expanding window, Table 14 suggests superiority of the AR-SARV model. Choosing a rolling window induces deterioration of the AR-SARV model. A QL loss of 4.1458 emerges for this model. This high QL loss additionally unveils instability in the variance predictions generated by the AR-SARV model[XIV]. As a result, the slight superiority of the AR-SARV model from the expanding window approach (QL loss of 2.1552) deteriorates drastically.

In contrast, the SR-SARV model worsens only marginally and still outperforms the MS-GARCH models by selecting the models according to the QL loss. In the expanding window prediction approach, the (QL loss is 2.1844), whereas in the rolling window approach the (QL loss is 2.2681). In conclusion, a rolling window approach only improves the variance prediction performance of MS-GARCH models and worsens the out-of-sample performance of SARV models. The instability of the AR-SARV specification is also noteworthy, implying that the formal conception of SARV models is crucial when predicting the conditional variance of Bitcoin log returns. The general finding is that, unlike MS-GARCH models, SR-SARV models

---

[XIV] The visualisation of the variance prediction of the AR-SARV model and the variance proxy shows that the AR-SARV model in combination with a rolling window is unsuitable. The visualisation is not reported here, but is generated by the corresponding R implementation.

(especially the AR-SARV specification) require a larger set of information $I_t$ to improve the variance prediction abilities.

## 5. Conclusions

Based on the obtained results, this research concludes that, in contrast to the commonly used AR(1) processes in the GARCH literature for modeling Bitcoin's volatility, a more thorough analysis suggests that ARMA processes of higher order may be a more suitable choice for future research. This finding holds particular significance in relation to the theoretical assumptions of MS-GARCH models. If researchers intend to employ MS-GARCH models with the aforementioned specifications, it is imperative to have a mean process that is free from serial correlation in order to obtain unbiased results. The estimation results of the ARMA model indicate that only higher-order processes are effective in eliminating all linear dependencies from the model residuals. Subsequently, MS-GARCH models were adapted to these residuals. The estimation of various MS-GARCH models indicates that the simple GARCH model is the most suitable model for describing the conditional variance of Bitcoin log returns. To be more precise, all information criteria (AIC, BIC, and HQ) consistently suggest that the simple GARCH model performs best in both volatility regimes. All simple GARCH processes within all regimes exhibit an extremely high level of persistence. While the ARCH and GARCH parameters of the simple GARCH processes recommended by the BIC are not significantly different, the AIC and HQ criteria select an MS-GARCH model where the simple GARCH processes differ only in terms of the composition of the ARCH and GARCH parameters. In this model, the standardized innovations are assumed to follow a skewed student-t distribution in the AIC- and HQ-preferred specification. However, the BIC favors a more parsimonious model, where the skewness and heavy-tail properties of the standardized innovation distributions are treated separately, with a student-t distribution in the first regime and a skewed normal distribution in the second regime.

The in-sample parameter estimation of SARV models exhibits a high degree of sensitivity to initial values, which can have notable consequences. The intricacies arise from the intricately designed weighted least squares target function, particularly when there are heteroscedastic innovations present in both the state and observation equations of the linear state space model. This intricacy can result in numerical optimization algorithms drifting during the estimation process, leading to the estimation of extreme values for certain SARV model parameters.

This sensitivity to initial values is a significant limitation of SARV models, as it can pose challenges during their implementation and the subsequent re-estimation of parameters for updates. It's worth noting that this sensitivity issue persists even when we separately estimate the mean and variance equations. It is reasonable to anticipate that simultaneous estimation of both the mean and variance equations within a state-space modeling framework could exacerbate these numerical estimation challenges, potentially leading to even more pronounced issues. Consequently, our recommendation to perform separate estimations for both equations proves to be more suitable, as it mitigates some of the numerical estimation complexities associated with SARV models. Thus, it is important to note that the in-sample estimation of SARV model parameters strongly relies on the specific target function being optimized and may also be influenced by the properties of the numerical optimization algorithm itself. Possible improvements could be achieved by using alternative optimization algorithms, such as the Broyden-Fletcher-Goldfarb-Shanno (BFGS) algorithms, employing a different target function, or reducing the number of model parameters through a model redesign. These enhancements have the potential to significantly enhance the predictive capabilities of SARV models.

The adjustment of the a-priori variance to obtain the a-posteriori variance, which represents the conditional variance in SARV models, depends on the specific SARV model specification. In the SR-SARV model, the Kalman gain is considered a dynamic and non-stationary process that adapts to changing market conditions, particularly in volatile phases. On the other hand, in the AR-SARV specification, the adjustment of the a-priori variance is less pronounced during volatile market phases. These differences in the behavior of the Kalman gain and variance adjustments reflect the varying dynamics of SARV models under different specifications and market conditions.

In the context of prediction, two distinct window approaches were employed: an expanding window and a rolling window with daily parameter re-estimation for the models. Interestingly, the rolling window approach yielded contrasting effects on the prediction performance of the models. While it improved the variance predictions of the MS-GARCH models, it had a detrimental impact on the predictions generated by the SARV models. Notably, this approach also exposed instability in the AR-SARV specification when applied within a rolling window framework.

Specifically, the prediction performance deteriorated significantly for the AR-SARV model, whereas the SR-SARV model experienced only a marginal decline. Across both window approaches, the SR-SARV model consistently outperformed the MS-GARCH models, as evidenced by lower QL losses.

Future research endeavors could explore the determination of the optimal window size for MS-GARCH models and assess whether SARV models, featuring various specifications for the conditional variance process, require a more extensive information set to enhance their variance prediction capabilities.

Overall, the results from the prediction section underscore that, based on the QL error measure, MS-GARCH models exhibit inferior predictive performance compared to the SR-SARV model. The latter consistently yielded smaller overall QL losses when evaluated under both rolling and expanding window approaches. While the MAE and MSE error measures exhibited small differences between SARV and MS-GARCH models, the QL loss emerged as the decisive criterion in this analysis, given its capacity to capture the predictive accuracy of the models effectively.

Investing in Bitcoin and other cryptocurrencies necessitates a robust risk management strategy. Existing research has cast doubts on the suitability of simple GARCH models for effective risk management due to their inability to handle non-stationary conditional variance processes. Recent studies propose the use of Markov-Switching GARCH models to address market inefficiencies, which can be attributed to factors such as manipulation, insider trading, and the actions of large cryptocurrency holders. Within this context, SARV models, particularly the AR and SR specifications, have emerged as promising alternatives, showcasing superior predictive capabilities when compared to MS-GARCH models.

This study's findings support the hypothesis that SARV models may excel in inefficient market conditions, thanks to their Kalman filter-based design. Furthermore, it is intriguing to note that, in certain scenarios, a highly persistent simple GARCH model might be sufficient for predicting the variance of Bitcoin log returns. This insight is grounded in the observation that GARCH-based variance processes within each regime can evolve independently and concurrently, resulting in more accurate variance predictions compared to the weighted approach that relies on filtered probabilities to calculate the overall conditional variance.

In light of these findings, this study recommends the adoption of an SR-SARV model as a foundational element within risk management frameworks. This recommendation is bolstered by the model's statistical sparsity and its ability to address the inefficiencies inherent in the cryptocurrency market.

**Bibliography**

**Antonopoulos, Andreas M. (2017)**: Mastering Bitcoin, 2. Ed., O'Reilly, Sebatopol, p. 1-14, 213-268.

**Ardia, David, Keven Bluteau and Maxime Rüede (2019)**: "Regime changes in Bitcoin GARCH volatility dynamics", in: *Finance Research Letters*, Vol. 29, p. 266-271.

**Ardia, David, Keven Bluteau, Kris Boudt, Leopoldo Catania and Denis-Alexandre Trottier (2016)**: "Markov-Switching GARCH Models in R: The MSGARCH Package", in: *Journal of Statistical Software*, Forthcoming, available online (accessed on 23th September 2019, 07:32): http://dx.doi.org/10.2139/ssrn.2845809.

**Bollerslev, Tim (2008)**: Glossary to ARCH (GARCH), CREATES Research Paper 2008-49, available online (accessed on 22th September 2019, 17:32): http://dx.doi.org/10.2139/ssrn.1263250.

**Bollerslev, Tim (1986)**: "Generalized Autoregressive Conditional Heteroscedasticity", in: *Journal of Econometrics*, Vol. 31 (3), p. 307-327.

**Bouri, Elie, Georges Azzi and Anne Haubo Dyhrberg (2017)**: "On the return-volatility relationship in the Bitcoin market around the price crash of 2013", in: *Economics*, Vol. 11 (2), p. 1-16.

**Box, George E. P., Gwilym M. Jenkins and Gregory C. Reinsel (2008)**: Time Series Analysis, 4th Ed., John Wiley & Sons, Hoboken, p. 193-231.

**Bradley, O. Brendan and Murad S. Taqqu (2003)**: "Financial Risk and Heavy Tails", in: Svetlozar T. Rachev (ed.), *Handbook of heavy Tailed Distributions in Finance*, North Holland, p. 35-103.

**Brownlees, Christian, Robert Engle and Bryan Kelly (2011)**: "A practical guide to volatility forecasting through calm and storm", in: *The Journal of Risk*, Vol. 14 (2), p. 3-22.

**Burniske, Chris and Adam White (2017)**: Bitcoin: Ringing the Bell for A New Asset Class, ARK INVEST + COINBASE research white papers, available online (accessed on 18th September 2019, 14:29): https://research.ark-invest.com/hubfs/1_Download_Files_ARK-Invest/White_Papers/Bitcoin-Ringing-The-Bell-For-A-New-Asset-Class.pdf.

**Caporale, Guglielmo Maria and Timur Zekokh (2019)**: "Modelling volatility of cryptocurrencies using Markov-Switching GARCH models", in: *Research in International Business and Finance*, Vol. 48, p. 143-155.

**Caporale, Guglielmo Maria, Gil-Alana and Alex Plastun (2018)**: "Persistence in the cryptocurrency market", in: *Research in International Business and Finance*, Vol. 46, p. 141-148.

**Caporale, Guglielmo Maria, Nikitas Pittis and Nicola Spagnolo (2003)**: "IGARCH models and structural breaks", in: *Applied Economics Letters*, Vol. 10 (12), p. 765-768.

**Casals, José, Alfredo Garcia-Hiernaux, Miguel Jerez, Sonia Sotoca, and A. Alexandre Trindade (2018)**: State-Space Methods for Time Series Analysis: Theory, Applications and Software, Chapman and Hall/CRC Press, Boca Raton, Chapters 4, 5 and 6.

**Chen, Shi, Cathy Yi-Hsuan Chen, Wolfgang Karl Härdle, T.M. Lee and Bobby Ong (2016)**: A first econometric analysis of the CRIX family, Economic Risk Berlin, SFB 649 Discussion Paper No. 2016-031, available online (accessed on 22th September 2019, 17:44): http://dx.doi.org/10.2139/ssrn.2832099.

**Chu, Jeffrey, Stephen Chan, Saralees Nadarajah and Joerg Osterrieder (2017)**: "GARCH Modelling of Cryptocurrencies", in: *Risk and Financial Management*, Vol. 10 (4), p. 1-15.

**Cont Rama (2007)**: "Volatility Clustering in Financial Markets: Empirical Facts and Agent–Based Models", in: Teyssiere, Gilles and Alan P. Kirman (eds.): *Long Memory in Economics*, Springer, Heidelberg, p. 289-310.

**Dickey, David A. and Wayne A. Fuller (1981)**: "Likelihood Ratio Statistics for Autoregressive Time Series with a Unit Root", in: *Econometrica*, Vol. 49 (4), p. 1057-1072.

**Dudek, Grzegorz, Piotr Fiszeder, Paweł Kobus and Witold Orzeszko (2024):** "Forecasting cryptocurrencies volatility using statistical and machine learning methods: A comparative study", in: *Applied Soft Computing*, Vol. 151, p 1-21.

**Dyhrberg, Anne Haubo (2016)**: "Hedging capabilities of bitcoin. Is it the virtual gold?", in: *Finance Research Letters*, Vol. 16, p. 139-144.

**Engle, Robert F. (1982)**: "Autoregressive Conditional Heteroscedasticity with Estimates of the Variance of United Kingdom Inflation.", *Econometrica*, Vol. 50 (4), p. 987–1007.

**Feng, Wenjun, Yiming Wang and Zhengjun Zhang (2017)**: "Informed trading in the Bitcoin market", in: *Finance Research Letters*, Vol. 26, p. 63-70.

**Fidelity Investments (2019)**: New Research From FIDELITY® Finds Institutional Investments in Digital Assets are Likely to Increase Over the Next Five Years, available online (accessed on 18th September 2019, 19:52): https://www.fidelity.com/bin-public/060_www_fidelity_com/documents/press-release/institutional-investments-in-digital-assets-050219.pdf.

**Fiedler, Salomon, Klaus-Jürgen Gern, Dennis Herle, Stefan Kooths, Ulrich Stolzenburg, Lucie Stoppok (2018)**: Virtual Currencies – Monetary Dialogue July 2018, In-Depth Analysis Requested by the ECON committee, available online (accessed on 11th October 2019, 20:31): http://www.europarl.europa.eu/committees/en/econ/monetary-dialogue.html?tab=2018.

**Fleming, Jeff and Chris Kirby (2003)**: "A Closer Look at the Relation between GARCH and Stochastic Autoregressive Volatility", in: *Journal of Financial Econometrics*, Vol. 1 (3), p. 365-419.

**Forney, G. David (1973)**: "The Viterbi Algorithm", in: *Proceedings of the IEEE*, Vol. 61 (3). p. 268-278.

**Francq Christian and Jean-Michel Zakoian (2011)**: GARCH Models: Structure, Statistical Inference and Financial Applications, John Wiley & Sons, Chichester, p. 234-340.

**Gandal, Neil, J. T. Hamrick, Tyler Moore and Tali Oberman (2018)**: "Price manipulation in the Bitcoin ecosystem", in: *Journal of Monetary Economics*, Vol. 95, p. 86-96.

**Glosten, Lawrence R., Ravi Jagannathan and David E. Runkle (1993)**: "On the Relation between the Expected Value and the Volatility of the Nominal Excess Return on Stocks", in: *The Journal of Finance*, Vol. 45 (5), p. 1779-1801.

**Griffin, John M. and Amin Shams (2018)**: Is Bitcoin Really Un-Tethered?, available online (accessed 16th June 2019, 20:41): http://dx.doi.org/10.2139/ssrn.3195066.

**Guo, Zi-Yi (2022)**: "Risk management of Bitcoin futures with GARCH models", in: *Finance Research Letters*, Vol. 45, p. 1-6.

**Haas, Markus, Stefan Mittnik and Marc S. Paolella (2004)**: "A New Approach to Markov-Switching GARCH Models", in: *Journal of Financial Econometrics*, Vol. 2 (4), p. 493-530.

**Hamilton, D. James (1989)**: "A New Approach to the Economic Analysis of Nonstationary Time Series and the Business Cycle", in: *Econometrica*, Vol. 57 (2), p. 357-384.

**Herle, Dennis (2018)**: Volatilitätsanalyse des Bitcoins, Kassel University.

**Jarque, M. Carlos and Anil K. Bera (1980)**: "Efficient tests for normality, homoscedasticity and serial independence of regression residuals", in: *Economic Letters*, Vol. 6 (3), p. 255-259.

**Katsiampa, Paraskevi (2017)**: "Volatility estimation for Bitcoin: A comparison of GARCH models", in: *Economics Letters*, Vol. 158, p. 3-6.

**Kurihara, Yutaka and Akio Fukushima (2018)**: "How Does Price of Bitcoin Volatility Change?", in: *International Research in Economics and Finance*, Vol. 2 (1), p. 8-14.

**Ljung, Greata M. and George E. P. Box (1978)**: "On a measure of lack of fit in time series model", in: *Biometrika*, Vol. 65 (2), p. 297-303.

**Nelder, John A. and Roger Mead (1965)**: "A Simplex Method for Function Minimization", in: *The Computer Journal*, Vol. 7 (4), p. 308-313.

**Nelson, Daniel B. (1991)**: "Conditional Heteroskedasticity in Asset Returns: A New Approach", in: *Econometrica*, Vol. 59 (2), p. 347-370.

**New York State Attorney General Office (2019)**: "Attorney General James Announces Court Order Against "Crypto" Currency Company Under Investigation For Fraud", Announcement dated 25th April 2019, available online (accessed on 10th October 2019, 23:52): https://ag.ny.gov/press-release/2019/attorney-general-james-announces-court-order-against-crypto-currency-company.

**Pesaran, M. Hashem and Allan Timmermann (2002)**: "Market timing and return prediction under model instability", in: *Journal of Empirical Finance*, Vol. 9(5), p. 495-510.

**Radovanov, Boris, Aleksandra Marcikic and Nebojsa Gvozdenovic (2018)**: "A Time Series Analysis of Four Major Cryptocurrencies", in: *Economics and Organization*, Vol. 15 (3), p. 271-278.

**Renault, Eric (2009)**: "Moment-Based Estimation of Stochastic Volatility Models", in: Andersen, Torben G., Richard A. Davis, Jens-Peter Kreiß and Thomas Mikosch (eds.): *Handbook of Financial Time Series*, Springer, Berlin, p. 269-312.

**Schwert, G. William (1990)**: "Stock Volatility and the Crash of '87", in: *Review of Financial Studies*, Vol. 3, p. 77-102.

**Shumway, Robert H. and David S. Stoffer (2017)**: Time Series Analysis and Its Applications – With R Examples, 4th Ed., Springer, Cham, p. 292-295.

**Trimborn, Simon and Wolfgang Karl Härdle (2016)**: CRIX an Index for Blockchain Based Currencies, Economic Risk Berlin, SFB 649 Discussion Paper 2016-021, available online (accessed on 03th October 2019, 13:26): http://dx.doi.org/10.2139/ssrn.2800928.

**Tsay, Ruey S. (2010)**: Analysis of Financial Time Series, 3rd Ed., John Wiley & Sons, Hoboken, p. 29-49.

**Urquhart, Andrew (2018)**: "What causes the attention of Bitcoin?", in: *Economic Letters*, Vol. 166, p. 40-44.

**Zakoian, Jean-Michel (1994)**: "Threshold heteroskedastic models", in: *Journal of Economic Dynamics and Control*, Vol. 18 (5), p. 931-955.